%% file: manuscript.tex
\documentclass[12pt]{article}

\usepackage[T1]{fontenc}                
\usepackage[utf8]{inputenc}             
\usepackage{crimson}                    
\usepackage{helvet}                     

\usepackage[strict,autostyle]{csquotes} 
\usepackage[USenglish]{babel}           
\usepackage{microtype}                  

\usepackage{abstract}                   
\usepackage{authblk}                    
\usepackage{booktabs}                   
\usepackage{caption}                    
\usepackage[final]{draftwatermark}      
\usepackage{geometry}                   
\usepackage{graphicx}                   
\usepackage{hyperref}                   
\usepackage{rotating}                   
\usepackage{setspace}                   
\usepackage{titlesec}                   
\usepackage{url}                        
\usepackage[numbers]{natbib}    
\usepackage{adjustbox}
\usepackage{multirow}
\usepackage{subfigure}

\usepackage{amsmath}
\usepackage{algorithm}
\usepackage[noend]{algpseudocode}

\usepackage{color}

\makeatletter
\def\BState{\State\hskip-\ALG@thistlm}
\makeatother


\newcommand{\myname}{Milan Janosov }
\newcommand{\myemail}{Corresponding author: rsin@itu.dk}
\newcommand{\myaffiliation}{Department of Network and Data Science\\Central European University, H-1051 Budapest, Hungary}
\newcommand{\paperdate}{\today}
\newcommand{\papertitle}{Success and luck in creative careers}
\newcommand{\paperkeywords}{success, dynamics of impact, creative careers, science of science}

\titleformat{\section}{\normalfont\sffamily\large\bfseries\color{black}}{\thesection.}{0.3em}{}
\titleformat{\subsection}{\normalfont\sffamily\small\bfseries\color{black}}{\thesubsection.}{0.3em}{}

\hypersetup{
	pdfauthor={\myname},
	pdftitle={\papertitle},
	pdfsubject={\papertitle},
	pdfkeywords={\paperkeywords},
	pdffitwindow=true,         
	breaklinks=true,           
	colorlinks=false,          
	pdfborder={0 0 0}          
}

\SetWatermarkText{DRAFT}
\SetWatermarkScale{1.3}
\SetWatermarkLightness{0.9}

\begin{document}

\title{\papertitle}
\date{\paperdate}
\author{\myname$^1$, Federico Battiston$^{1}$, Roberta Sinatra$^{2,3,4}$\footnote{\myemail}}

\affil[]{\small
					$1$) \myaffiliation\\
					$2$) Department of Computer Science, IT University of Copenhagen, 2300 Copenhagen, Denmark  \\
					$3$) ISI Foundation, 10126 Torino, Italy \\
					$4$) Complexity Science Hub Vienna, 1080 Vienna, Austria\\
				}

\maketitle


\begin{abstract}

Luck is  considered to be a crucial ingredient to achieve impact in all creative domains, despite their diversity. For instance, in science, the movie industry, music, and art, the occurrence of the highest impact work and of a hot streak within a creative career are very difficult to predict. Are there domains that are more prone to luck than others? Here, we provide new insights on the role of randomness in impact in creative careers in two ways: (i) we systematically untangle luck and individual ability to generate impact in the movie, music, and book industries, and in science, and compare the luck factor between these fields; (ii) we show the limited predictive power of collaboration networks to predict career hits. Taken together, our analysis suggests that luck consistently affects career impact across all considered sectors and improves our understanding in pinpointing the key elements in the prediction of success.

\textbf{Keywords:} \paperkeywords

\vspace{0.5cm}
\end{abstract}


\section{Introduction}

Research in developmental psychology has studied careers of prominent artists and scientists for decades, advocating the importance of chance for the successful unfolding of careers in various creative domains~\cite{lehman1953age, campbell1960blind, simonton1984creative, simonton1988age}. In recent years, the availability of big databases on scientific publications~\cite{wos} and artistic records, from books to movies~\cite{spitz2014measuring, yucesoy2018success, williams2019quantifying}, has made it possible to test a number of previously suggested hypotheses on a large scale. For instance, in  previous work~\cite{sinatra2016quantifying,liu2018hot} the analysis of thousands of creative careers has shown that the biggest hit of an individual occurs randomly within an individual's career, a finding named the equal-odds-rule~\cite{simonton1984creative}. This rule explains the variability in the occurrence of creative individuals' best hit. Yet, career hits are not only the results of luck but also of other individual and team properties~\cite{guimera2005team,uzzi2013atypical,wang2013quantifying,lee2015creativity,zagovora2018increases,fortunato2018science,jadidi2018gender}. While previous literature suggests that luck and individual ability are both necessary to excel in art and science charts~\cite{galton1889hereditary,flugel1964hundred,petersen2011quantitative, pluchino2018talent, pluchino2019exploring}, a quantification of the role of luck across different creative domains is still lacking. In which creative fields are individuals more likely to go from rags to riches and vice-versa? Does the position of an individual in a network predict the occurrence of a hit?

In this work, we quantify luck fluctuations in impact across creative careers from movies, music, literature, and science, and create a framework to compare the broad observed differences in impact~\cite{radicchi2008universality, yucesoy2018success}. Do these random fluctuations have the same magnitude across careers? To address this question we build on the mathematical framework known as the $Q$-model proposed in Ref.~\cite{sinatra2016quantifying} to untangle the impact into two components, one encoding fluctuations that can be interpreted as luck, and another depending only on the individual. We show that this model is consistent with the classical test theory~\cite{crocker1986introduction}, also known as the true score theory~\cite{lord1965strong}, stating that the measured value of a certain measurable attribute consists of the sum of its true -- error-free -- score, and a stochastic error term. We find that the value of such \textit{randomness} varies depending on the creative fields. By comparing this stochastic term to the typical impact score associated with each artist and scientist, we identify creative domains where the impact of single creative products are the hardest to predict and fluctuate the most within individual careers. The high importance of luck to achieve success in creative careers is confirmed by the lack of power of the collaboration networks to predict the best hit of an individual. To carry out these analyses, we rely on a large-scale data set covering more than four million individuals from c. 1902 up until 2017. 

The outline of this paper is the following. First, we test the validity of the requirements of the {\it Q}-model proposed in Ref.~\cite{sinatra2016quantifying}. Second, we use the {\it Q}-model impact decomposition method to factor impact in creative careers.  Third, we apply the classical test theory to quantify the role of luck within each field and discuss the observed differences across fields.  Finally, we construct the collaboration network within each domain and  compare the time of the best hit of creative individuals to the time at which they reach their highest score in network centrality.

\begin{figure}
\centering
\includegraphics[width=0.95\textwidth]{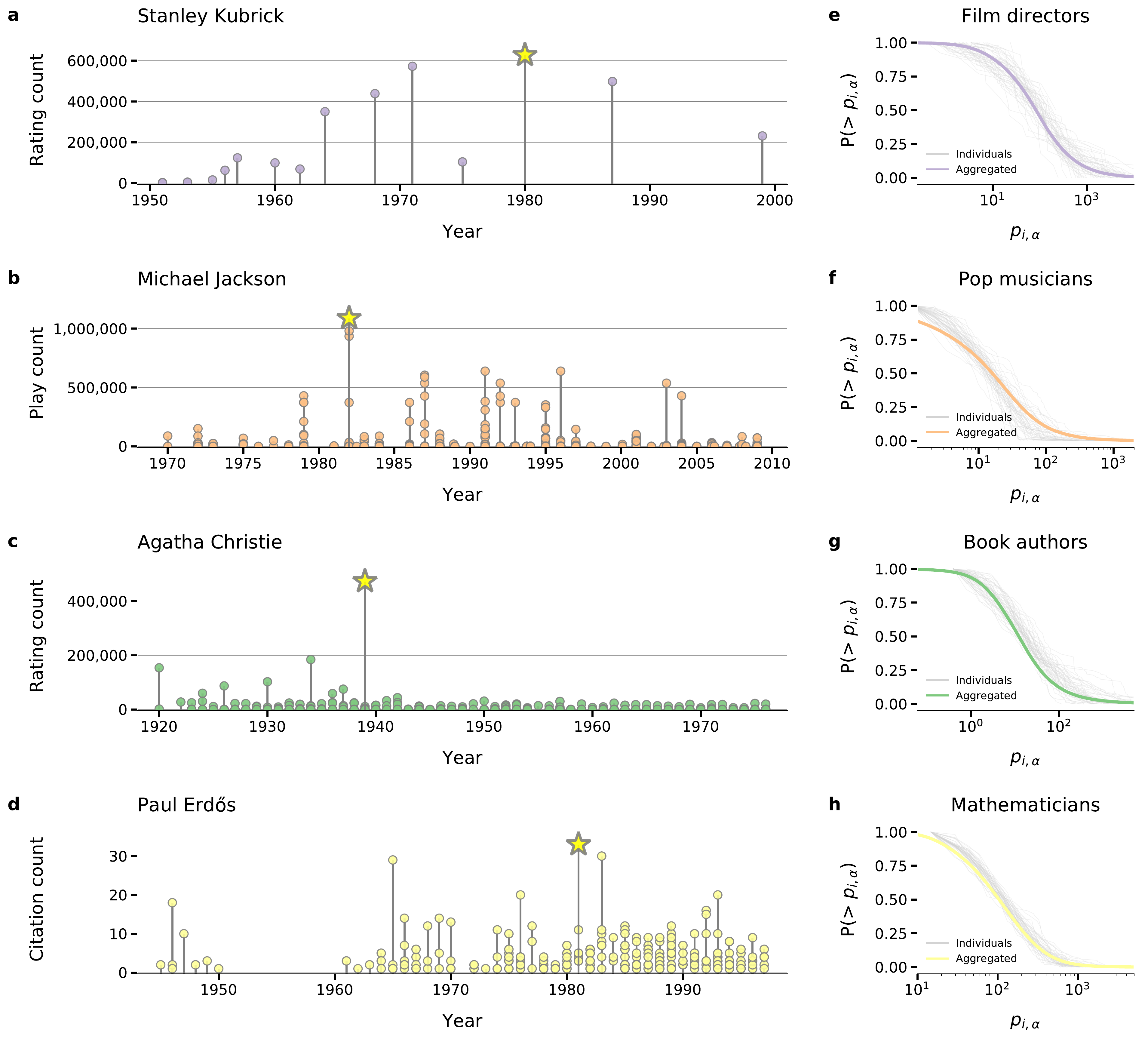}
\caption{{\bf Career examples and rescaled impact distributions in four creative domains.}  \textbf{a}, The career trajectory of Stanley Kubrick. On the horizontal axis, we show the release year of his movies, while on the vertical axis we show the impact of each movie, captured by the number of ratings received from IMDb users. \textbf{b}, The career of Michael Jackson. We show the release year of his songs and the song impact captured by the total play count on the music provider LastFM. \textbf{c}, The career of Agatha Christie. We report her books publication dates and the book impact, captured by the number of ratings they received on Goodreads. \textbf{d}, Publication history of Paul Erdős, mathematician and graph theorist, based on his record in the Web of Science database. The paper impact is measured by the number of total citations 10 years after publication.  
\textbf{e-h}, The rescaled cumulative impact distribution $P(p_{i,\alpha})$, where $p_{i,\alpha}=S_{i,\alpha}/Q_{i}$ for \textbf{e}, movies of  directors, \textbf{f}, tracks of musicians  active in pop music,  \textbf{g}, books of authors, and \textbf{h}, papers of mathematicians. The figures show that when we rescale the impact value of each product of an individual by his/her $Q$ parameter, their distribution collapses onto roughly the same aggregated curve, marked by continuous colored lines. The distribution of 50 randomly chosen individuals is visualized by light grey lines. }
\label{fig:figure1}
\end{figure}

\begin{figure}[htp]
\centering
\includegraphics[width=0.95\textwidth]{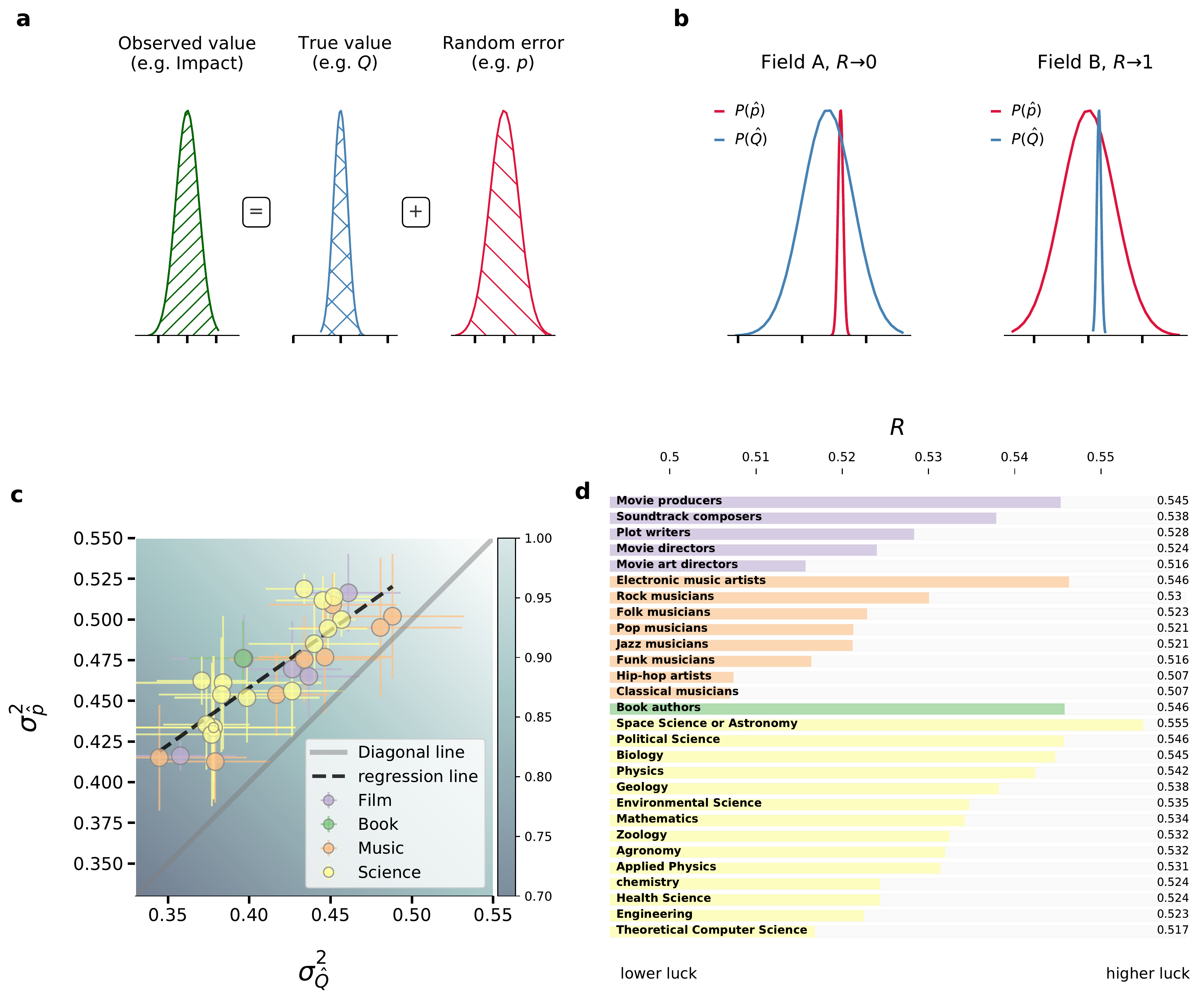} 
\caption{\textbf{Fluctuations of impact, luck and $Q$}. {\bf a,} According to the classical test theory, the normal distribution of an observed variable (green in the example) can be composed as the sum of the distributions of the true score (blue) and the error term (red). {\bf b,} Distribution of $\hat{p}$ and of $\hat{Q}$ for two different, fictional fields. In Field A the distribution of $\hat{p}$ has a low variance compared to $\hat{Q}$, therefore randomness has a negligible role ($R \to 0$). Field B exhibits the opposite behavior, with a narrow $\hat{Q}$ and broad $\hat{p}$ distribution meaning that the individual's luck dominates impact ($R \to 1$).  {\bf c,} We show the studied 28 creative fields on the $\left(\sigma^2_{\hat{Q}}, \sigma^2_{\hat{p}}\right)$ plane, marking fields from different data sets with different colors. We denoted a fitted line by continuous black line and added the diagonal as continuous grey line as a reference. The gradient-coloring of the background changes in a diagonal direction, illustrating that the points being on the same off-diagonal line have the same ${\rm Var}  \hat{S}$ (SI Section 2.3.1) {\bf d,} The table shows the values of the $R$ randomness index for the different fields.} 
\label{fig:figure2}
\end{figure}

\begin{figure}[htp]
\raggedright
\includegraphics[width=0.95\textwidth]{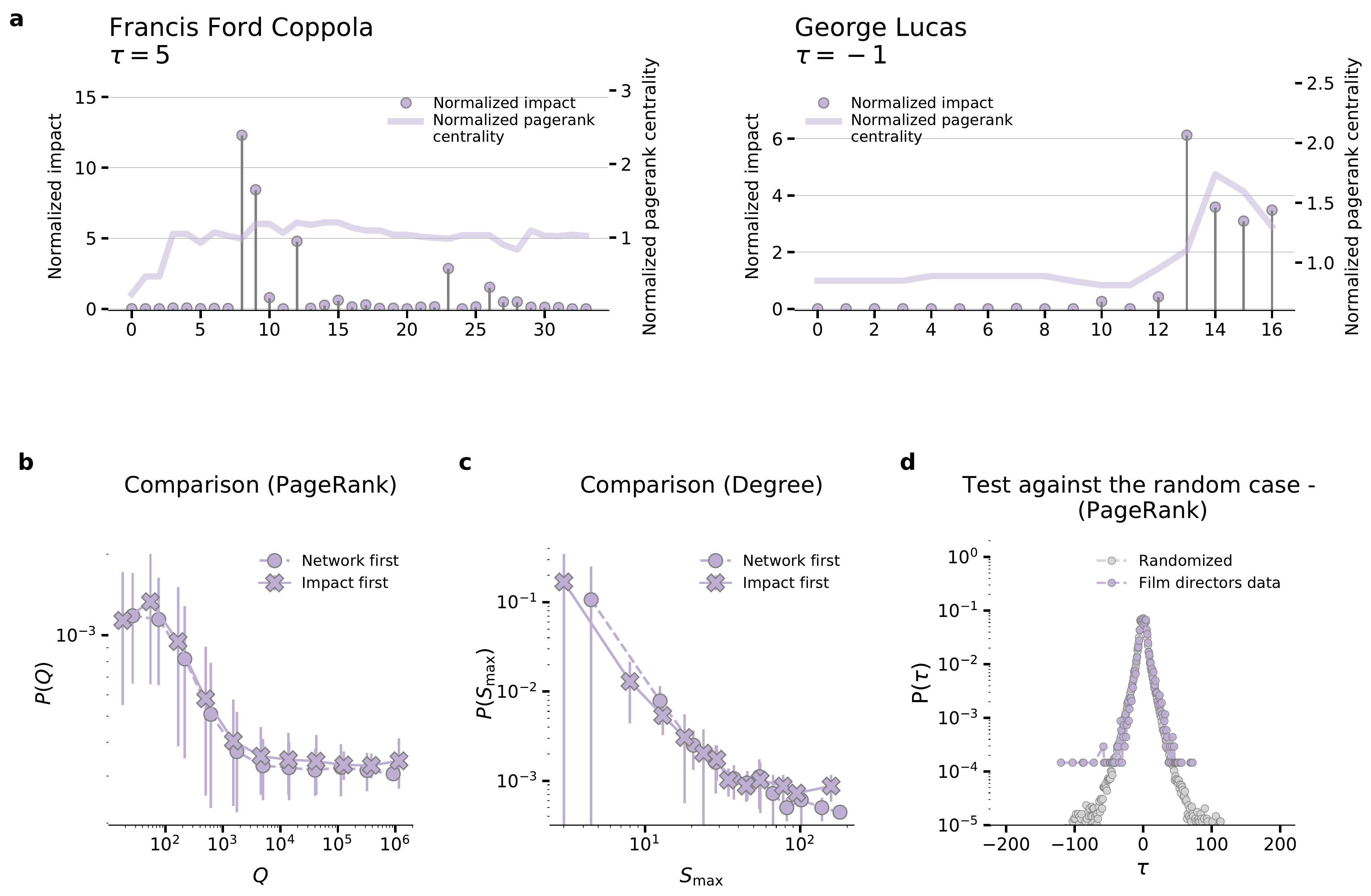}
\caption{{\bf Network position and timing of biggest hit for movie directors.} {\bf a,} We report the impact $S$ of Francis Ford Coppola  illustrating the case when PageRank centrality peaks first ($\tau = 5$), followed by the impact, while the example of George Lucas  illustrates the opposite behavior ($\tau = +1$), where a peak in impact is followed by the peak in networking. The figure then shows a comparison between the groups of film directors for whom success peaks before (colored continuous lines) and after (colored dashed lines) their network peak. For movie directors, {\bf b,} their network position based on their PageRank and their success measured by their $Q$ parameters, {\bf c,} their network position based on their degree and their highest impact, the binned (15 bins) distributions of the two groups of do not show a significant difference based on the Kolmogorov-Smirnov test ($d_{\rm degree} = 0.02$, $d_{\rm PageRank} = 0.02$, $p <0.0001$). {\bf d,} Shows the distribution of the shift parameter $\tau$ between the directors' network centrality (PageRank) and impact time series, coloring the distribution corresponding to the original data by orange, and to the randomized data by orange (KS test $d=0.27$, $p < 0.01$.}
\label{fig:figure3}
\end{figure}



\section{Data}

We compiled four data sets and individual careers across the movie, music, and book industries, and across scientific fields, covering overall 28 different types of creative domains:
\begin{enumerate}
    \item We mined the Internet Movie Database (IMDb~\cite{imdb}) and compiled a data set of  803,013  individuals in the movie industry working as movie directors, producers, art directors, soundtrack composers, and scriptwriters, altogether contributing to 1,297,275 movies. 
    \item By using the Discogs~\cite{discogs, hartnett2015discogs} and LastFM~\cite{lastfm} platforms, we constructed a database of 379,366 musicians released 31,841,981 songs in the genres of electronic, rock, pop, funk, folk, jazz, hip-hop, and classical music. 
    \item We extracted data from Goodreads~\cite{goodreads} and built a data set containing information about 2,069,891 book authors and  6,604,144  books.
    \item We used the Web of Science database~\cite{wos} to reconstruct the scientific careers of 1,204,688 scientists from the fields of chemistry, mathematics, physics, applied physics, space science and astronomy, zoology, geology, agronomy, engineering, theoretical computer science, biology, environmental science, political science, and health science, altogether authoring approximately 87,4 million papers.
\end{enumerate}
See further details about the data sets and data collection in  SI Section S1.1.

To measure the impact of movies, songs, books, and articles, we use their cumulated impact on large audiences, as captured by the rating counts for movies and books, the play counts for songs, and the number of citations received within the first ten years after publication for scientific papers~\cite{garfield1979citation} (SI Section S1.2). The existence of these cumulative impact measures in all data sets allows us to reconstruct individual careers consistently across domains by building the historical time series of each person. In Figure \ref{fig:figure1}a-d we illustrate career examples in the four different databases: movie director Stanley Kubrick, pop singer Michael Jackson, writer Agatha Christie, and mathematician Paul Erdős. Alternative impact measures, like the average rating for movies or books, or rescaled citations for papers~\cite{radicchi2011rescaling}, highly correlate with the cumulative measures used here, indicating that the impact patterns do not depend on the chosen measure (see SI Section S1.3 for details).


\section{The $Q$-model: decomposing luck and individual ability in impact}
\label{Sec:results_modeling}

Kubrick's highest impact movie was released 30 years after his career start, while Michael Jackson had his biggest hit earlier in his career. These anecdotal examples suggest that a career's biggest hit can occur at any time. Indeed, a rigorous analysis of our data sets indicates that any work in a career has an equal chance to be the highest impact work, following the so-called \textit{random-impact-rule}, consistently with what previously found for large data sets of artists and scientists~\cite{sinatra2016quantifying,liu2018hot} (SI Section S2.1 for this replication analysis). The magnitude of a career impact is not random though: individual impact distributions differ broadly from each other. These broad differences are reproduced and explained by the $Q$-model~\cite{sinatra2016quantifying}, a mechanistic stochastic model (SI Section S2.2). According to this model, the impact $S_{i, \alpha}$ of a work $\alpha$ created by an individual $i$ can be decomposed as the product of two independent factors $S_{i, \alpha} = Q_i p_{i, \alpha}$, where $Q_i$ is an individual variable, depending only on individual $i$, and $p_{i, \alpha}$ is a stochastic variable, independently drawn for every work from a field-specific distribution. The values of $Q_i$ and $p_{i, \alpha}$ are obtained by maximizing a likelihood function which takes as input all the impact $S$ of all products of all creative careers in a given field ~\cite{sinatra2016quantifying,  vasarhelyi2018optimized}.

Under the main assumption that the covariance $\sigma^2_{QN}$ between the distributions of productivity $N$ and parameter $Q$ is negligible compared to the variance of the $p$ and $N$ distributions -- an assumption that we verify and validate in SI Section S2.3 -- we can write a simple approximated formula for $Q_i$:
\begin{eqnarray}
Q_i &=& e^{\left\langle \log{S_{i,\alpha}} \right\rangle - \mu_p },  \label{eq:qapprox}
\end{eqnarray}
where $\mu_p$ is the mean of the $p$ distribution within a given field. Eq.~(\ref{eq:qapprox}) indicates that the exponent of $Q_i$ is the average of the order of magnitude of the impact of $i$'s works, minus a constant equal for all individuals in a field. To establish whether the $Q$-model reproduces the individual impact distributions in our data sets, we first check the hypothesis that both $S$ and $N$ follow a log-normal distributions  (SI Section S2.3). We then estimate the parameters associated with the distributions of $p$ and $Q$, finding that within each creative domain both $Q_i$ and $p_{i, \alpha}$ are also log-normally distributed (SI Subsection S2.3.3). 

The negligible measured covariance $\sigma^2_{pN}$ and $\sigma^2_{pQ}$ predict that the individual rescaled impact, $p_{i, \alpha}=S_{i, \alpha} / Q_i $, should follow a universal distribution, independent of $Q_i$. We use this prediction to validate the model in our data sets: we measure the distribution $p_{i, \alpha}=S_{i, \alpha} / Q_i $ and show that it collapses roughly on a single curve for different careers (Figures \ref{fig:figure1}e-h). Since this rescaled distribution is independent of individual variables like $N_i$ and $Q_i$, we can interpret $p$ as a ``luck factor" driving impact~\cite{sinatra2016quantifying}. Finally, we compare the data with the the scaling of the highest impact work with productivity as predicted by the $Q$-model, and show that the $Q$-model gives significantly better results than the random model (SI Section S2.4).

A single high impact work in a career is not sufficient to have a high $Q_i$; rather an individual needs to perform consistently well throughout her career. For instance, the movie director with the highest $Q_i$, Christopher Nolan, has a $Q_i=1719.3$, due to his many high impact movies like ``Inception" or ``Interstellar".
In contrast, one-hit wonders, who achieved fame with a single song or movie, and whose success was neither anticipated nor repeated throughout their career with many high impact works, are typically characterized by lower values of $Q_i$. An example is Michael Curtiz (1886–1962), director of the all-time classic Casablanca, who has only a modest $Q_i= 4.8$ as he did not direct any other movies with outstanding impact. In this case, the large impact of their career's biggest hit is explained by a lucky draw of a high $p$, rather than being due to the individual ability to consistently produce work of high impact, encoded in a high $Q$. Taken together, the $Q$-model well reproduces the career impact of individuals in our data sets.


\section{From the $Q$-model to classical test theory to compare luck across different domains}
\label{Sec:results_luckandskill}

Here we introduce a quantitative approach, based on the $Q$-model, to compare the fluctuations in luck and variations in the typical impact across different creative fields. Recalling the impact decomposition $S_{i, \alpha} = Q_i p_{i, \alpha}$ presented in in Section \ref{Sec:results_modeling}, we can write:
\begin{eqnarray}
\hat{S}_{i, \alpha} &=& \hat{Q}_{i} + \hat{p}_{i, \alpha} \label{eq:eq2},
\end{eqnarray}
where $\hat{S}_{i, \alpha} = \log S_{i, \alpha}$,  $\hat{Q_i} = \log Q_i$  and  $\hat{p}_{i, \alpha} = \log p_{i, \alpha}$. Because $p$ and $Q$ are log-normally distributed (SI subsection S2.3.3), $\hat{p}$  and $\hat{Q}$ are normally distributed. In addition,  the covariance $\sigma^2_{pQ} \approx 0$, then $\sigma^2_{\hat{p}\hat{Q}} \approx 0$. Therefore, Eq. (\ref{eq:eq2}) takes the form proposed by classical test theory~\cite{lord1965strong, kristof1974estimation, kline2005psychological, kean2014item, mauboussin2010untangling, mauboussin2012success} for decomposing the measured value of a certain quantity. Namely, according to this theory, the measurable value of an observed attribute, in this case $\hat{S}$, can be decomposed as the sum of two uncorrelated variables both following normal distributions. One of these two variables encodes the true score of the quantity, in this case, $\hat{Q}$, and the other variable encodes a random error term, $\hat{p}$ (Figure \ref{fig:figure2}a). The two normal distributions of the variables $\hat{Q_i}$ and $\hat{p}_{i, \alpha}$ are in line with previous studies, suggesting that individual variables like skill and talent, and global ones such as luck, are typically normally distributed~\cite{stewart1983distribution, crocker1986introduction, allen2001introduction,  mauboussin2010untangling, mauboussin2012success, pluchino2018talent}. 

Building on Eq. (\ref{eq:eq2}), on the properties of normal distributions and on the measured properties of the $Q$ and $p$ variables in our data sets, we can express the variance of $\hat{S}_{i, \alpha}$, $\sigma^2_{\hat{S}}$, as:
\begin{eqnarray}
\sigma^2_{\hat{S}} &=& \sigma^2_{\hat{Q}} + \sigma^2_{\hat{p}},
\label{eq:eq22}
\end{eqnarray}
where $\sigma^2_{\hat{p}}$ and $\sigma^2_{\hat{Q}}$ are the variance of the distributions of $\hat{p}$ and $\hat{Q}$, respectively. 
This decomposition allows us to measure the relative importance of the luck component compared to the individual component in determining impact. Building on previous work~\cite{mauboussin2010untangling, mauboussin2012success}, we define the {\it randomness} index $R$ capturing the share of luck in the overall impact variance as:
\begin{eqnarray}
R &=& \frac{\sigma^2_{\hat{p}}}{\sigma^2_{\hat{S}}} \label{eq:eq3}.
\end{eqnarray}
When individuals in a domain have a similar ability, captured by a narrow $\hat{Q_i}$ distribution, differences in impact are mainly driven by luck, and we have that $R \to 1$. In contrast, when {\it p} has a low variance compared to {\it S}, then $R \to 0$,  and luck plays only a small role. This index allows us to compare the role of  randomness across 28 different creative fields (Figure \ref{fig:figure2}b).

\section{Randomness in creative careers}
In which creative domains are inequalities driven more by luck than by individual ability? Using the $Q$-model, we measure  $\sigma^2_{\hat{Q}}$ and $\sigma^2_{\hat{p}}$ for 28 types of creative careers in the movie, music, and book industries, and in science (Figure \ref{fig:figure2}c). We also report the linear regression between $\sigma^2_{\hat{p}}$ and $\sigma^2_{\hat{Q}}$ (black dashed line on Figure \ref{fig:figure2}c). This figure offers a number of findings. First, we observe that all the fields are placed above the diagonal line ($\sigma^2_{\hat{p}} > \sigma^2_{\hat{Q}}$), indicating that within each domain fluctuations in luck are broader than those in the typical career impact of individuals.  Second, we do not observe any domain-specific clustering on the $\left(\sigma^2_{\hat{Q}}, \sigma^2_{\hat{p}}\right)$ plane, which suggests that the studied domains do not differ from each other for luck. Third, we report that the linear regression has a slope lower than one; therefore, it intercepts the diagonal for high $\sigma^2_{\hat{Q}}$. Because $\sigma^2_{\hat{S}}=\sigma^2_{\hat{p}} + \sigma^2_{\hat{Q}}$ and the regression slope is equal to the ratio $\sigma^2_{\hat{p}} / \sigma^2_{\hat{Q}}$, a value smaller than 1 indicates that as $\sigma^2_{\hat{S}}$ increases (illustrated by the shading on Figure \ref{fig:figure2}c), the value of $\sigma^2_{\hat{Q}}$ increases faster than $\sigma^2_{\hat{p}}$. Hence large fluctuations in impact are dominated by large fluctuations in individual ability, captured by $Q$, rather than fluctuations in luck.

Next, we measure the randomness index $R$ of Eq. (\ref{eq:eq3}) to compare the characteristics of career success across domains (Figure \ref{fig:figure2}d). We find that on the one hand, within the movie industry, producers' careers are the most driven by luck, followed by composers. On the other hand, being an art director is associated with the lowest $R$ index, suggesting that high impact as an art director happens less by chance than in other careers within the movie industry. It is also interesting to compare the randomness index of scriptwriters ($R = 0.528$) and book authors ($R = 0.546$), due to the apparent similar nature of these two creative careers. The value of the indexes show that writing for the movie industry is less driven by luck than in the book industry. In music, classical and hip-hop are the most robust against luck fluctuations with the lowest randomness index of our data set, $R = 0.507$. This could be explained by classical music being more dependent on skills, experience, and musical training. Regarding hip-hop music, we could speculate that being largely an underground genre, it is less exposed to the rich-gets-richer effect and leaves more space for rising fresh talents. In contrast, the most popular genres, namely electronic music ($R=0.546$) and rock music ($R=0.530$), are on the other side of the spectrum with the highest $R$. These two genres contain the largest number of one-hit wonder careers; therefore impact has more pronounced fluctuations. Regarding science, we also find a large spectrum of randomness, with space science and astronomy ($R=0.555$) and political science ($R=0.546$), at one extreme for the highest $R$-index fields, and theoretical computer science ($R=0.517$) and engineering ($R=0.523$) being among the most robust fields against luck fluctuations.


\section{Lack of predictive power of the collaboration network}
In the previous sections, we have analyzed the randomness and magnitude of impact focusing on individual careers. However, a movie, a song or a paper is rarely the result of the work of only one individual. Therefore next we ask: Can collaborations between individuals improve our ability to predict the magnitude of success and the occurrence of career big hits? Previous research suggests that scientific career success and network position can be connected \cite{figg2006scientific,hsu2011correlation, radicchi2012science, sarigol2014predicting, zagovora2018increases, janosov2019elites}. 

We reconstruct the temporal aggregated network of movie directors, pop musicians, and mathematicians to study the relationship between their network positions and impact. We use a yearly time
resolution. In this network, each individual is represented by a node, and the strength of the connection between nodes at year $T$ is the Jaccard-index of the works of the two nodes, that is the number of works the two both individuals collaborated on, divided by the total number of works they contributed to until year $T$. Based on this definition, the final aggregated collaboration network of movie directors consists of 8,091,208 links between 184,220 people active between 1927-2017 (giant connected component only). In the pop music network, we have 52,366 musicians active between 1926-2017 connected by 8,232,349 links, while in mathematics, we have 94,755 links between 27,401 mathematicians between 1944 and 2016. 
 
For each individual, we measure her degree centrality, PageRank centrality, and clustering coefficient in the aggregated network at the time she has produced a work. We then create individual time-series for each of these network measures, where time points correspond to the works in the individual career. Finally we study these network-based time-series together with the evolution of individual impact over a career. Our hypothesis is that the dynamics of the network position and the dynamics of impact are correlated over time, however with a delay of $\tau$. We measure $\tau$ by shifting the network time-series with respect to the impact time-series, and choose the value for which we obtain the maximum correlation between the time-series  (see Figure \ref{fig:figure3} SI Section S3).   

By analyzing the time-series of movie directors, pop musicians, and mathematicians, we find that there are two groups of individuals: those for whom the network measures peak before the highest impact work occurs, and those for whom the peak occurs after.
For example, the director Francis Ford Coppola ($\tau = 5$) belongs to the first category, while George Lucas ($\tau$ = -1) to the second (Figure \ref{fig:figure3}a). However, there are no discernible differences between these two groups when we measure impact: the two groups have similar distributions of the $Q$-parameter (Figure \ref{fig:figure2}b) and of the magnitude of the highest success withing a career (Figure \ref{fig:figure2}c). 

Given the indistinguishable nature of impact in these two groups, we ask whether the observed shift $\tau$ is different from that obtained from reshuffled time-series, where time correlations are canceled. We measure the distribution of the delay parameter $\tau$, and compared it to the distribution of a randomized data set in which the time-series are randomly reshuffled. The two distributions are closely overlapping, confirmed by the double-sided Kolmogorov-Smirnov test (Figure \ref{fig:figure3}d, details about the KS test in SI Section S3). Taken together, the collaboration network among individuals does not improve our ability to predict the timing of the biggest hit, suggesting that chance has much higher importance than the collaboration network to determine the timing of the biggest hit within a career.


\section*{Conclusion}

In this work, we provided a framework to understand and quantify the role of randomness in the success of creative fields across different domains. To understand the emergence of high-impact creative works, we built large-scale data sets and investigated thousands of careers from the movie, book, and music industries, and from science. We built on an existing model, known as $Q$-model, to decompose the impact of the individual creative works into two independent components, one expressing the ability of an individual to have consistently high or low typical impact, captured by the \textit{$Q$-parameter}, and one associated to random fluctuations, capturing the role of \textit{luck}. We also cast the model into the framework of classical test theory, which aims to disentangle the true score of a variable from noisy fluctuations.
Using this framework, we found that on average fluctuations in impact of single creative works are more influenced by luck than by individual ability. However, we conclude that the fluctuations in the individual parameter are more pronounced for fields with large fluctuations in impact. The extrapolated linear trend between fluctuations in individual parameter and in luck predicts that when impact fluctuations become large ($\sigma^2_{\hat{Q}}\approx0.6$), the fluctuations in individual parameter become larger than the random ones. In this ideal, not observed case, the fluctuations in impact would be mainly due to individual differences. Moreover, we found that sub-disciplines within different domains cannot be clustered according to the relative magnitude of these fluctuations. The absence of clustering suggests the magnitude of luck is not a distinctive feature of domains.
We introduced a synthetic randomness index, defined as the relative ratio of the variance of the random component to that of success, and investigated its score across different domains. We found that the randomness index varies in a relatively narrow range, despite the differences in typical impact. This further confirms the lack of distinct typical scales of random fluctuations associated with the four different domains investigated in the paper. Finally, in this narrow range of randomness, we found that the careers with highest luck are those of movie producers, electronic music artists, book authors, and scientists working in the fields of space science, and political science. On the other hand, randomness has the lowest influence on hip-hop and classical music, theoretical computer science, and movie art directors.
Finally, we also studied the temporal relationship between success and centrality in the collaboration network for  movie directors, pop musicians, and mathematicians as a case study. For each individual, we compared the temporal evolution of their network centrality to the evolution of their impact. We found that these two are correlated, yet with a delay. We computed these delay parameters and found two distinct classes of creative careers regardless to their creative domain. Individuals belonging to the first group produce their big hit first, and become well-connected in the network only after the occurrence of the hit, while people falling into the second category first build favorable connections, and produce their big hit afterwards. However, we found no correlation between individual impact and group the individual belongs to. We also showed that the delay between the impact and the network time-series follows the same distribution as randomized data.

Future studies could further untangle  the individual $Q$-parameter and pinpoint what $Q$ means, for example in terms of access to resources or early career steps. Also the variable $p$, interpreted here as luck, could contain more information than just randomness, if further data is incorporated in the analysis. Nevertheless, its universal distribution across careers suggests that this information is homogeneously distributed among individuals.


\section*{Competing interests}
The authors declare that they have no competing financial or non-financial interests.

\section*{Author's contributions}
R.S. conceived the study. M.J., F.B., and R.S. collaboratively designed the study, and drafted, revised, and edited the manuscript. M.J. analyzed the data and ran all numerical analyses.

\section*{Acknowledgements}
Special thanks to Em\H{o}ke-\'Agnes Horv\'ath, J\'anos Kert\'esz, Federico Musciotto, Rossano Schifanella, Michael Szell, G\'abor V\'as\'arhelyi for their valuable suggestions.

\section*{Funding}
M.J. and R.S. acknowledge support from Air Force Office of Scientific Research grant FA9550-15-1-0364. The authors declare that they have no competing financial interests.

\section*{Availability of data and materials}
The processed data files and scripts to reproduce the results presented on the figures are available here:  \url{https://github.com/milanjanosov/Success-and-randomness-in-creative-careers}


\begin{small}
\setlength{\bibsep}{0.00cm plus 0.05cm} 
\bibliographystyle{unsrt}
\bibliography{refs}
\end{small}

\clearpage

\section{Supplementary information}
\input{SupplementaryInformation}

\end{document}

%% file: SupplementaryInformation.tex
\renewcommand{\thesection}{S.I. \arabic{section}}
\renewcommand{\thetable}{S.I. \arabic{table}}
\renewcommand{\thefigure}{S.I.\arabic{figure}}
\setcounter{figure}{0}    
\setcounter{table}{0}


\subsection{Data}


\subsubsection{Data sets}
\label{subsectionSI:datasets}

Our research was based on four different data sources which were collected during the period of June-August 2017. 
\textbf{IMDb dataset.} We collected information on individuals active in the movie industry based on the Internet Movie Database (IMDb~\cite{imdb}). To this scope we first used the Advanced Title Search\footnote{{http://www.imdb.com/search/title}} function and sent multiple queries to obtain the list of all movie identifiers which received a vote from at least  one user. Using the list of unique movie identifiers ($\sim$1.3 million) we downloaded the HTML source code of each movie's site.  After processing all the raw HTML files about the movies we filtered out $\sim$0.8 million distinct names being listed as director, producer, scriptwriter, composer, and art-director, and created the career files by associating each movie to the corresponding individuals, for each professions separately (e.g. directors, producers). We attached the six different success measures present in the database: average rating, rating count, metascore~\cite{meta}, gross revenue to each career, the number of user and critic reviews, to each career and constructed the individuals' career trajectories as time series of these quantities.

\textbf{Discogs and LastFM dataset.} To cover individuals active in the music industry we relied on Discogs\footnote{{https://www.discogs.com/search/}}~\cite{discogs}, a crowd-sourced music discography website. Via its search functionality, we listed all the master releases from the genres of rock, pop, electronic, folk, funk, hip-hop, classical, and jazz music to obtain a comprehensive list of $\sim$0.4 million artists combined. After crawling their  discographies based on their unique identifiers from Discogs and parsing them into tracklists, we used the API of LastFM, a music providing service (www.lastfm.com~\cite{lastfm}), to extract the play counts used as impact measures. For each song we queried the complete tracklist of the artists, and kept only those which had been played at least once. This way, we obtained a dataset consisting of $\sim$31 million songs. Then we combined the timestamped discography and the song -- play count datasets to reconstruct the musicians' careers for each genre.

\textbf{Goodreads dataset.} We gathered data about book authors using Goodreads (www.goodreads.com,~\cite{goodreads}), a social network site for readers by crawling the HTML website of the profile of $\sim$2.1 millions individuals authored $\sim$6.6 millions books. By extracting information from the authors' biography profiles we built their career trajectories. Goodreads provides three different way for measuring impact: the average users' ratings of a book, the total number of ratings, and the number of editions a book has, from which we used the rating count for further analysis. 

\textbf{Web of Science dataset.} We used the Web of Science~\cite{wos} database to reconstruct the careers of scientists from 15 scientific disciplines: Agronomy, Applied Physics, Biology, Chemistry, Engineering, Environmental Science, Geology, Health Science, Mathematics, Physics, Political Science, Space Science Or Astronomy, Theoretical Computer Science, Zoology.
In total, we analyse the careers of 1.55 million scientists, who authored 87.4 million papers. Each paper has been associated with the number of citations received.The career of a scientist consists of her publication record and the citation impact of each paper.

\vspace{0.7cm}
After collecting these data sets, to limit the analysis to careers with sustained productivity, we set a filtering thresholds to 10 movies and papers for individuals with movie and scientific careers (except art-directors, for whom it was 20), 50 books for authors, and 80 songs for musicians. The relatively high threshold for artists active in the music industry is due to releases usually containing multiple songs.


\subsubsection{Measuring success in artistic domains}
\label{subsectionSI:meassuccess}

\begin{figure}[bp]
\centering
\includegraphics[width=0.75\textwidth]{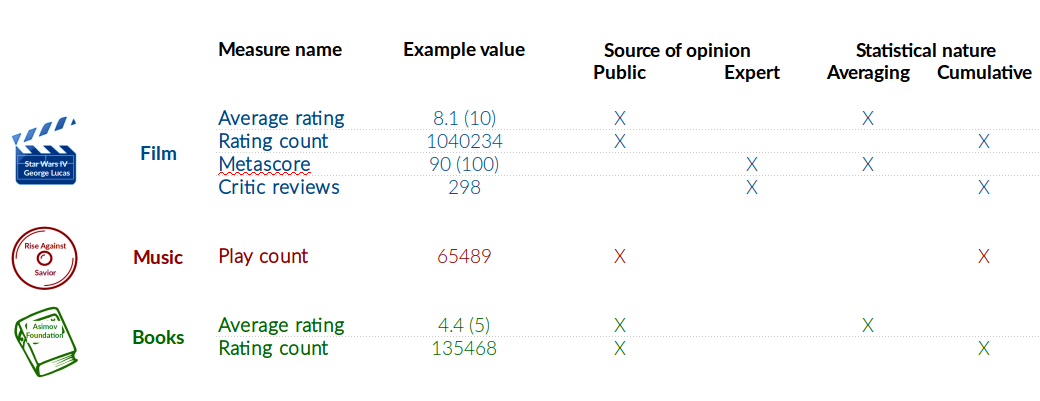}
\caption{{\bf Taxonomy of success measures.} Success measures associated to movies, songs and books differ based on their definition (public opinion or experts' evaluation), and their statistical nature (averaged or cumulated over time), defining four distinct categories of success measures. Measures associated with all possible four combinations are present for movies (IMDb~\cite{imdb}). Songs (Discogs and Lastfm~\cite{discogs, lastfm}) are only associated with a public opinion-based cumulative measure. Data about books (Goodreads~\cite{goodreads}) provide only public opinion based measures.  } 
\label{fig:SF0}
\end{figure}

Our research premise is that success is a social phenomenon, and as such we aim to capture ``a community’s reactions to the performance of the individuals''~\cite{yucesoy2016untangling,barabasi2018formula}. For this reason, the movies, songs, books, and scientific papers in our database are associated with measures of success of different nature based on their social context. On one hand, there are success measures that are based on the evaluation of experts of the field, who have supposedly more insights on the underlying performance associated with the artistic product. On the other hand, success measures based on the opinion of the general public have larger statistics. However, they are also more likely to be biased by external factors, such as the rich-gets-richer phenomenon or the peer-effect~\cite{muchnik2013social}. From a statistical perspective, success measures can be either obtained as an average of responses over time or as the result of cumulative activities through time (Figure~\ref{fig:SF0}). We based our analysis on the cumulative measures since these are the only ones present in all different available data sets. This also allowed us to adopt existing techniques and methodologies previously used for the study of paper and scientific careers, which all capture success through different metrics based on the accumulation of citations .


\subsubsection{Correlations between different success measures}
\label{subsectionSI:correlations}

\begin{figure}[tb]
\centering
\includegraphics[width=1.0\textwidth]{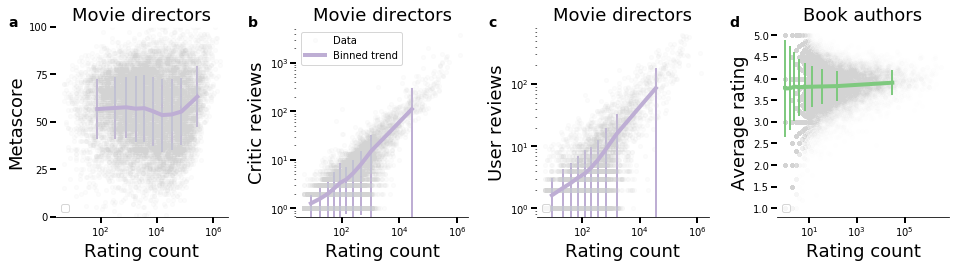}
\caption{\label{fig:SF2}{\bf Correlations between different success measures.} The correlations between movies'  rating counts and their \textbf{a}, Metascores (Spearman's rank correlation coefficient $r_S   \approx 0.151$); \textbf{b}, the number of critic reviews ($r_S\approx0.839$); and \textbf{c}, the number of user reviews ($r_S\approx0.641$) for the case of movie directors. Subfigure \textbf{d}, shows the low correlation between the number of ratings and the average rating a book receives on Goodreads ($r_S \approx 0.022$). In the scatter plots, each dot represents a product (i.e. a movie or a book), while the continuous lines show the percentile-binned trends using 10 bins.}
\end{figure}

Two of our data sets, covering movie and books, contain more than one type of success measures. Here we compare them by computing the correlations between pairs of measures of different kind. We find that different cumulative measures show high correlations with each other (see Fig. \ref{fig:SF2}b-c), indicating that results are robust to the choice of the specific cumulative measure. Average measures, like Metascore (Fig. \ref{fig:SF2}a) or average rating Fig. \ref{fig:SF2}d), do not correlate well with cumulative measures, indicating a different process generating these measures. Since these average measures have a broad distribution, and previous literature offers methods and finding mainly about cumulative measures, we opted to use cumulative measures.


\subsection{Q-model}
\label{sectionSI:qmodel}


\subsubsection{Testing the random impact rule}
\label{subsectionSI:rir}

\begin{figure}[hbt]
\centering
\includegraphics[width=1.00\textwidth]{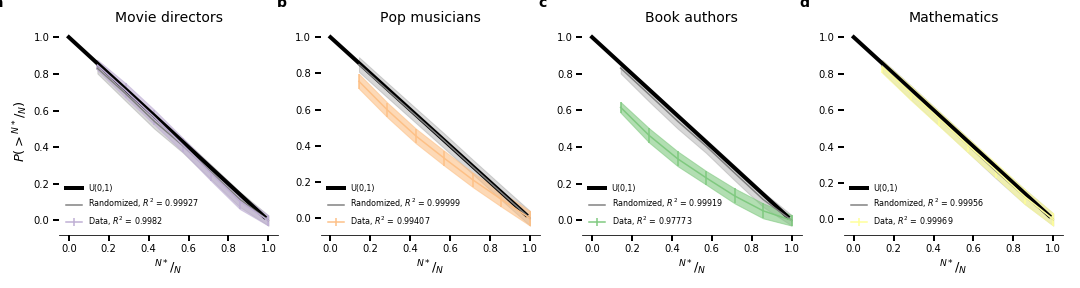}
\caption{{\bf Validating the random impact rule.} We validate the random impact rule by comparing the observed $P(N^*/N)$ to distribution obtained for randomized careers and to the theoretical CDF of a uniform distribution.}
\label{fig:SF6}
\end{figure}

The random impact rule states that the chronological rank of the best product ($N^*$) over a career with a length of $N$ is uniformly randomly distributed across a large sample of careers, meaning that the probability distribution $P(N^*/N)$ is well approximated by a uniform $U(0,1)$ distribution, as prior work has already shown for scientific fields~\cite{sinatra2016quantifying} and other creative domains~\cite{liu2018hot}. To test this hypothesis in our multiple creative domains, we compared the observed success cumulative distribution function (CDF) $P(> N^*/N)$ with both the CDF of the theoretical $U(0,1)$ distribution, and the CDF in a set of synthetic careers. In the synthetic careers, we randomly reshuffled the products, making sure that $N^*$ takes a uniformly randomly assigned position over the career. To obtain statistically reliable results, we repeated this randomization 100 times. We quantified the goodness of the fit by computing the $R^2$ deviation and the Kolmogorov--Smirnov distance of the original and the randomized data from the theoretical null model, i.e. the cumulative distribution function of the $U(0,1)$ uniform distribution. Results for four examples from each domain are shown on Figure \ref{fig:SF6}. The goodness of the fit for all the studied professions is measured by the $R^2$ value comparing the data to the $U(0,1)$. These results are summarized in Table \ref{tab:random}.

\begin{table}[]
\scriptsize
\begin{tabular}{l|l}
\textbf{Field}                 & \textbf{$R^2$}                       \\
\hline
\hline
Funk musicians                            & 0.95236                              \\
Rock musicians                          & 0.95565                              \\
Jazz   musicians                         & 0.96488                              \\
Folk musicians                           & 0.96508                              \\
Classical musicians                       & 0.96694                              \\
Movie art directors                   & 0.97207                              \\
Book authors                         & 0.97773                              \\
Pop musicians                             & 0.99407                              \\
Electronic music artists                        & 0.99553                              \\
Movie directors                       & 0.9982                               \\
Soundtrack composers                       & 0.99923                              \\
Movie producer                       & 0.99941                              \\
Theoretical Computer Science & 0.99954                              \\
Agronomy                       & 0.99955                              \\
Script writers                         & 0.99957                              \\
Space Science or Astronomy  & 0.99959                              \\
Zoology                        & 0.9996                               \\
Health Science                & 0.99962                              \\
Chemistry                      & 0.99963                              \\
Biology                        & 0.99967                              \\
Geology                        & 0.99967                              \\
Mathematics                    & 0.99969                              \\
Environmental  science         & 0.99969                              \\
Physics                        & 0.99972                              \\
Engineering                    & 0.99973                              \\
Political Science             & 0.99973                              \\
Applied Physics               & 0.99979 \\
\hline
\end{tabular}

\caption{
\vspace{0.1cm}
{\bf Validating the random impact rule.} Goodness of fit of the random impact rule measured with the $R^2$ value across the studied data sets.}
\label{tab:random}
\end{table}

\clearpage


\pagebreak
\subsubsection{Q-model}
\label{subsectionSI:qmodel}
The {\it Q}-model, proposed in~\cite{sinatra2016quantifying}, assumes that when the distribution of the impact of scientific papers can be described by log-normal functions, then the impact can be expressed as the trivariate log-normal distribution of three variables: \textit{(i)} the productivity of the individuals (e.g. the number of papers they publish, $N$), \textit{(ii)} an individual based parameter  only depending on the individual's prior works' success, $Q$), and \textit{(iii)} a random parameter representing outer factors ($p$). By transforming these variables to the logarithmic space ($\hat{N} = \log{N}$, $\hat{Q} = \log{Q}$, $\hat{p} = \log{p}$), the impact $P(\hat{S})$ distribution reads:
\begin{eqnarray}
\label{trivariate}
P(\hat{S}) &=& P(\hat{p}, \hat{Q}, \hat{N}) = \frac{1}{\sqrt{(2 \pi)^3 }} \exp \bigg( - \frac12 ({\bf X} - \mu)^T \Sigma^{-1} ({\bf X} -\mu)  \bigg),
\end{eqnarray}
where ${\bf X} = (\hat{p}, \hat{Q}, \hat{N})$, $\mu = (\mu_N, \mu_p, \mu_Q)$ is the average vector, and $\Sigma$ the covariance matrix: 
\vspace{0.1cm}
$\Sigma$ =
$\left( {\begin{array}{ccc}
 \sigma^2_p         &  \sigma_{p,Q}     &    \sigma_{p,N}  \\
\sigma_{p,Q}  &  \sigma^2_Q           &     \sigma_{Q,N}   \\
\sigma_{p,N}  &  \sigma_{Q,N}           &     \sigma^2_N   \\
  \end{array} } \right)$.
\vspace{0.1cm}
If the cross terms $\sigma_{p,Q}$ and $\sigma_{p,N}$ are close to zero, then the distribution of $p$ does not depend on variables capturing individual careers. In this case a number of simplications can be made, and the impact rescaled by the individual parameter $Q$ collapses on the same distribution for all individuals. 
To obtain the covariance matrix of the trivariate log-normal distribution of Eq. \ref{trivariate}, we fit the theoretical distribution to the data by using  CMA-ES~\cite{hansen1996adapting, vasarhelyi2018optimized}  (Covariance Matrix Adaptation Evolution Strategy. Evolution strategies), from which we obtained the parameters in Table \ref{tab:ST1}. The shown results are consistent with the reported findings about scientific careers in~\cite{sinatra2016quantifying}.

\begin{table}[]
\scriptsize
\begin{tabular}{llllllllll}
\textbf{Field}                          & \textbf{$\mu_N$} &
\textbf{$\mu_p$} & \textbf{$\mu_Q$} & $\sigma_N$ & $\sigma_Q$ & $\sigma_p$ & $\sigma_{pQ}$ & $\sigma_{pN}$ & $\sigma_{QN}$ \\
\hline
\hline
\textbf{Agronomy}                       & 4.467            & 2.148            & 4.147            & 1.711      & 0.452      & 0.514      & -0.069        & 0.025         & -0.005        \\
\textbf{Applied Physics}               & 4.521            & 1.224            & 2.938            & 1.823      & 0.399      & 0.452      & -0.046        & -0.015        & 0.012         \\
\textbf{Movie art directors}                  & 4.214            & 4.557            & 3.877            & 1.606      & 0.437      & 0.465      & -0.047        & 0.030         & 0.037         \\
\textbf{Biology}                        & 4.305            & 2.334            & 4.078            & 1.867      & 0.434      & 0.519      & -0.072        & -0.074        & -0.027        \\
\textbf{Books authors}                  & 3.514            & 2.868            & 5.571            & 1.682      & 0.396      & 0.476      & -0.054        & -0.014        & -0.062        \\
\textbf{Chemistry}                      & 4.366            & 2.462            & 4.586            & 1.629      & 0.440      & 0.485      & -0.059        & -0.132        & 0.083         \\
\textbf{Classical musicians}                & 5.313            & 4.762            & 5.896            & 1.873      & 0.488      & 0.502      & -0.072        & 0.024         & 0.003         \\
\textbf{Soundtrack composer}                       & 3.687            & 3.935            & 5.312            & 1.634      & 0.358      & 0.416      & -0.030        & 0.005         & -0.059        \\
\textbf{Movie directors}                       & 4.174            & 4.673            & 4.992            & 1.783      & 0.426      & 0.469      & -0.053        & 0.097         & -0.017        \\
\textbf{Electronic music artists}                        & 5.144            & 4.531            & 3.761            & 1.833      & 0.345      & 0.415      & -0.035        & 0.026         & 0.038         \\
\textbf{Engineering}                    & 4.976            & 2.170            & 3.768            & 1.601      & 0.457      & 0.500      & -0.063        & 0.017         & 0.046         \\
\textbf{Environmental Science}         & 3.489            & 2.112            & 4.194            & 1.517      & 0.445      & 0.512      & -0.063        & 0.034         & -0.025        \\
\textbf{Folk musicians}                           & 5.246            & 4.071            & 6.260            & 1.744      & 0.434      & 0.475      & -0.054        & -0.026        & -0.060        \\
\textbf{Funk musicians}                           & 5.608            & 3.612            & 6.081            & 1.656      & 0.447      & 0.477      & -0.058        & -0.076        & 0.011         \\
\textbf{Geology}                        & 4.592            & 4.749            & 4.441            & 1.802      & 0.374      & 0.435      & -0.037        & 0.071         & -0.056        \\
\textbf{Health Science}                & 4.114            & 3.986            & 3.513            & 1.702      & 0.448      & 0.494      & -0.061        & 0.048         & -0.023        \\
\textbf{Hip-hop artists}                         & 5.354            & 4.703            & 5.433            & 1.472      & 0.481      & 0.495      & -0.068        & -0.072        & -0.078        \\
\textbf{Jazz musicians}                           & 4.380            & 3.340            & 5.337            & 1.695      & 0.379      & 0.413      & -0.030        & -0.022        & -0.099        \\
\textbf{Mathematics}                    & 4.557            & 4.662            & 4.212            & 1.647      & 0.378      & 0.434      & -0.040        & -0.030        & -0.102        \\
\textbf{Physics}                        & 4.552            & 1.760            & 3.396            & 1.571      & 0.383      & 0.454      & -0.046        & 0.055         & 0.037         \\
\textbf{Political Science}             & 4.723            & 2.889            & 4.041            & 1.807      & 0.384      & 0.461      & -0.055        & -0.002        & 0.010         \\
\textbf{Pop musicians}                            & 6.071            & 2.553            & 4.421            & 1.911      & 0.417      & 0.454      & -0.046        & -0.024        & 0.018         \\
\textbf{Movie producers}                       & 3.823            & 4.424            & 4.010            & 1.499      & 0.397      & 0.476      & -0.052        & -0.081        & -0.019        \\
\textbf{Psychology}                     & 4.565            & 4.696            & 4.514            & 1.000      & 0.737      & 2.013      & 0.220         & 0.009         & -0.012        \\
\textbf{Rock musicians}                           & 5.518            & 3.858            & 3.643            & 1.572      & 0.451      & 0.509      & -0.069        & -0.017        & 0.046         \\
\textbf{Space Science Astronomy}  & 5.231            & 2.164            & 2.523            & 1.672      & 0.371      & 0.462      & -0.046        & -0.056        & 0.094         \\
\textbf{Theoretical Computer Science} & 4.058            & 4.781            & 4.216            & 1.799      & 0.426      & 0.456      & -0.045        & -0.006        & 0.042         \\
\textbf{Script writers}                         & 3.024            & 2.944            & 4.128            & 1.620      & 0.461      & 0.516      & -0.073        & -0.005        & -0.073        \\
\textbf{Zoology}                        & 4.058            & 3.655            & 3.523            & 1.723      & 0.377      & 0.429      & -0.042        & -0.028        & -0.048       \\
\hline
\end{tabular}
\vspace{0.1cm}
  \caption{{\bf Optimization results.} The table reports the parameters of the $P(N)$, $P(Q)$, and $P(p)$ distributions obtained my evolutionary optimization for all the studied fields.}\label{tab:ST1}
\end{table}



\subsubsection{Requirements of the {\it Q}-model}
\label{subsectionSI:requirements}
To apply the Q-model to a creative domain, the data has to fulfill a number of requirements. First, the observed impact distribution should follow a log-normal distribution, as shown in Section  \ref{subsubsectionSI:impactdistr}. Second, the random-impact-rule should hold (proven in Section \ref{subsectionSI:rir}). Third, $p$ should be uncorrelated from $Q$, and $N$ (their pairwise correlations should be negligible compared to their variances), which is shown in Table \ref{tab:ST1}. Finally, the distributions of $p$, $Q$, and $N$ is log-normal which we show in the following.


\paragraph{Fitting the impact distributions}
\label{subsubsectionSI:impactdistr}

\begin{figure}[htp]
\centering
\includegraphics[width=1.0\textwidth]{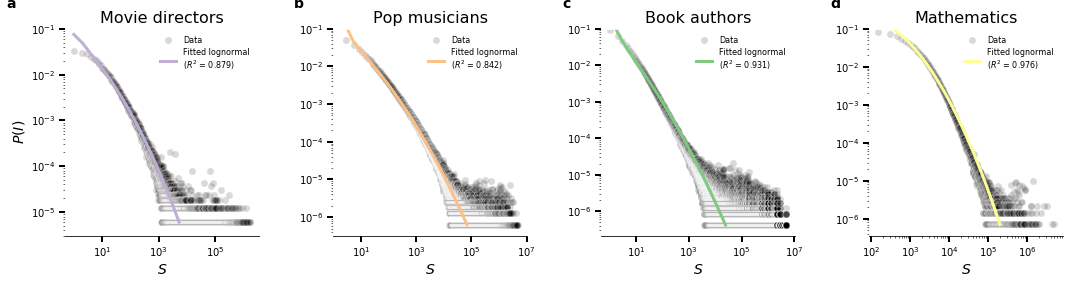}
\caption{\label{fig:SF4}{\bf Log-normal fitting of the impact distributions.} We report the impact distribution of one kind of career in each of the four considered data sets, namely movie directors, pop musicians, book authors, and mathematicians. The distribution is in light grey, the fitted curve is a colored continuous line. We measure the goodness of the fit by the coefficient of determination}
\includegraphics[width=1.0\textwidth]{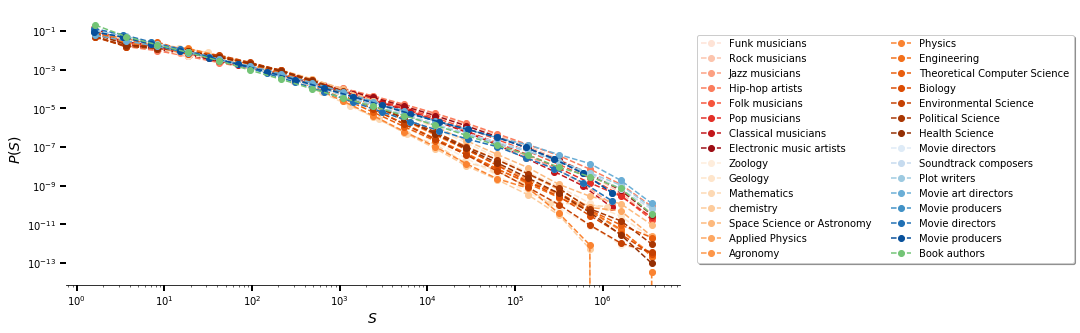}
\caption{\label{fig:SF4c}{\bf Re-scaled impact distributions.}  The min-max re-scaled impact distributions of the different fields.}
\end{figure}

To model the distribution of the success measure on the different fields -- rating count for movies and books,  play count for songs, and citations for scientific papers -- we assumed a log-normal shape and fitted the cumulative distribution function of the data (Examples from each data set are on Figure \ref{fig:SF4}).  We quantified the goodness of the fit by computing $R^2$ values, whose values are reported for all the fields in Table \ref{tab:impact}
\pagebreak

\begin{table}[]
\scriptsize
\begin{tabular}{l|l}
\textbf{Field}                 & \textbf{$R^2$}                       \\
\hline
\hline
Funk musicians                          & 0.95236                              \\
Rock  musicians                          & 0.95565                              \\
Jazz musicians                           & 0.96488                              \\
Folk musicians                           & 0.96508                              \\
Classical musicians                      & 0.96694                              \\
Movie art directors                   & 0.97207                              \\
Book authors                          & 0.97773                              \\
Pop musicians                            & 0.99407                              \\
Electronic music artists                        & 0.99553                              \\
Movie director                       & 0.9982                               \\
Movie composers                       & 0.99923                              \\
Movie producers                       & 0.99941                              \\
Theoretical Computer Science & 0.99954                              \\
Agronomy                       & 0.99955                              \\
Script writer                         & 0.99957                              \\
Space Science Astronomy  & 0.99959                              \\
Zoology                        & 0.9996                               \\
Health Science                & 0.99962                              \\
Chemistry                      & 0.99963                              \\
Biology                        & 0.99967                              \\
Geology                        & 0.99967                              \\
Mathematics                    & 0.99969                              \\
Environmental Science         & 0.99969                              \\
Physics                        & 0.99972                              \\
Engineering                    & 0.99973                              \\
Political Science             & 0.99973                              \\
Applied Physics               & 0.99979 \\
\hline
\end{tabular}
\vspace{0.1cm}
\caption{{\bf Goodness of fit of fitting log-normal curves onto the observed impact distributions.} The values of coefficient of determination ($R^2$) describing the goodness of fit of the log normal function on the observed impact distributions for all the studied careers. }
\label{tab:impact}
\end{table}

Since different fields reach a different audience, the impact of creative products across domains spans different ranges. In order to compare the decompositions of impacts across the $Q$ and $p$ components for several fields, beforehand we apply a min-max scaling to the measured impacts. This transforms $P(S_a)$, the impact distribution of field $a$, in the following way:
\begin{eqnarray}
P(S_a)  \quad \to \quad \frac{P(S_a)  - \min(P(S_a)} {\max(P(S_a)) - \min(P(S_a))}  \cdot \max(P(S_c)),
\end{eqnarray}
where $P(S_c)$ denotes the distribution of all the fields combined. The re-scaled impact distributions of the different fields are visualized in Figure \ref{fig:SF4c}.

\begin{figure}
\centering
\includegraphics[width=1.0\textwidth]{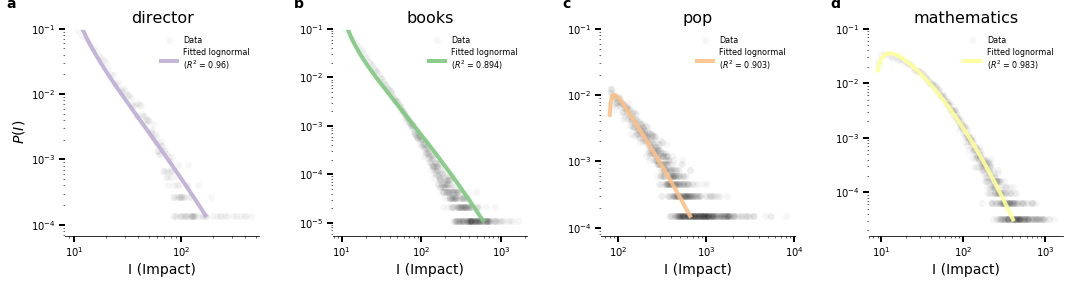}
\caption{\label{fig:career}{\bf The career length distributions fitted by log-normal curves of four selected fields.} We fit the productivity distribution (number of creative products each individual produced) by log-normal curves and characterized the goodness of fit by computing their $R^2$ values.}
\end{figure}

\paragraph{Career length distributions}
\label{subsubsectionSI:careerlength}
Figure \ref{fig:career} shows the log-normal distributions fitted on the career distributions for four selected, representative fields. The goodness of the fit is summarized in Table \ref{tab:length}.

\begin{table}[]
\scriptsize
\begin{tabular}{l|l}
\textbf{Field}                 & \textbf{$R^2$}                     \\
\hline
\hline
Jazz  musicians                         & 0.869                              \\
Script writers                        & 0.877                              \\
Hip-hop artists                         & 0.882                              \\
Book authors                        & 0.894                              \\
Pop  musicians                           & 0.903                              \\
Movie producers                       & 0.916                              \\
Movie art directors                   & 0.922                              \\
Folk musicians                           & 0.923                              \\
Electronic music artists                        & 0.947                              \\
Rock musicians                          & 0.948                              \\
Movie directors                       & 0.96                               \\
Space Science or Astronomy  & 0.965                              \\
Environmental Science         & 0.968                              \\
Movie composer                       & 0.974                              \\
Zoology                        & 0.976                              \\
Applied Physics               & 0.982                              \\
Geology                        & 0.982                              \\
Theoretical Computer Science & 0.982                              \\
Mathematics                    & 0.983                              \\
Biology                        & 0.984                              \\
Political Science             & 0.984                              \\
Engineering                    & 0.986                              \\
Agronomy                       & 0.988                              \\
Chemistry                      & 0.989                              \\
Physics                        & 0.989                              \\
health\_science                & 0.991 \\
\hline
\end{tabular}
\vspace{0.1cm}
\caption{{\bf The career length distributions fitted by lognormal curves.} We fit the productivity distribution (number of creative products each individual produced) by log-normal curves and characterized the goodness of fit by computing their $R^2$ values.}
\label{tab:length}
\end{table}

\paragraph{$P(Q)$ and $P(p)$}
\label{subsubsectionSI:pQdistribution}

\begin{figure}[htp]
\centering
\includegraphics[width=1.00\textwidth]{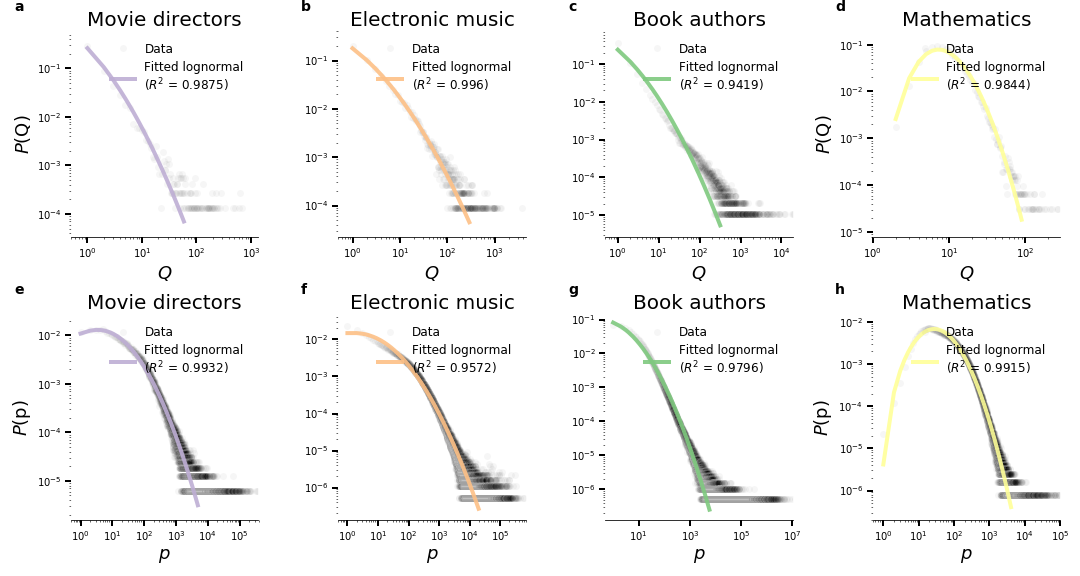}
\caption{{\bf The distributions of  $Q$ and  $p$ in different creative domains.} Based on the validation of $Q$-model in the studied fields, we are able to decompose the success of the works in the product of two parameters: the individual-based $Q$ parameter, which only depends on the career trajectory of the individual and is a unique constant for everyone, and $p$, a probabilistic parameter which is a random number drawn from the same distribution for all the products on the same field.  Figures illustrate how the values of the $Q$ and $p$ parameters (grey scatter plot and colored binned trend) are distributed in the different  fields and how well they compare to the lognormal model.}
\label{fig:SF8}
\end{figure}

The distributions $P(Q)$ and $P(p)$ described by log-normal functions, as illustrated by the fitted graphs on four representative fields on Figure \ref{fig:SF8}, and the results being summarized in Table \ref{tab:pq}

\begin{table}[]
\scriptsize
\begin{tabular}{l|l|l}
\textbf{Field}                 & \textbf{$R^2_p$} & \textbf{$R^2_Q$}                    \\
\hline
\hline
Classical musicians                      & 0.9444           & 0.9933                              \\
Applied Physics               & 0.9572           & 0.9654                              \\
Pop musicians                            & 0.9742           & 0.9844                              \\
Engineering                    & 0.9776           & 0.9793                              \\
Book authors                         & 0.9796           & 0.9796                              \\
Movie composers                       & 0.9801           & 0.9822                              \\
Zoology                        & 0.9807           & 0.9206                              \\
Agronomy                       & 0.9808           & 0.9419                              \\
Folk musicians                          & 0.981            & 0.973                               \\
Movie producers                       & 0.9817           & 0.974                               \\
Movie writer                         & 0.9841           & 0.9751                              \\
Mathematics                    & 0.9856           & 0.9778                              \\
Electronic music artists                        & 0.9857           & 0.9813                              \\
Biology                        & 0.9884           & 0.9822                              \\
Geology                        & 0.9885           & 0.9824                              \\
Jazz musicians                          & 0.989            & 0.9844                              \\
Political Science             & 0.9913           & 0.9844                              \\
Movie director                       & 0.9914           & 0.9875                              \\
Hip-hop artists                         & 0.9915           & 0.9875                              \\
Funk  musicians                          & 0.9916           & 0.988                               \\
Theoretical Computer Science & 0.9918           & 0.9886                              \\
Space Science or Astronomy  & 0.9928           & 0.9922                              \\
Environmental Science         & 0.9932           & 0.9933                              \\
Rock musicians                           & 0.9934           & 0.9952                              \\
Chemistry                      & 0.9945           & 0.996                               \\
Physics                        & 0.9961           & 0.9972                              \\
Movie art director                   & 0.9989           & 0.9994 \\   \hline
\end{tabular}
\vspace{0.1cm}
\caption{{\bf The distributions of $Q$ and  $p$ in different creative fields.} The goodness of the fit in all the studied fields for both the $p$ and $Q$ distributions characterized by the $R^2$ value.}
\label{tab:pq}
\end{table}


\subsubsection{Comparison to the data}
\label{subsectionSI:comparison}

We tested a simple model for the success of creative products in individual careers, based on the random impact rule, by generating sets of careers on each field based on the random impact rule. We then compared the highest impact for work in the synthetic careers to that of the observed data. To ensure that the set of synthetic careers is directly comparable to data, we constructed them by randomly reshuffling the time events of the careers found in the data, then repeated this random shuffling 100 times and averaged them to minimize the level of noise.

\begin{figure}[htp]
\centering
\includegraphics[width=1.00\textwidth]{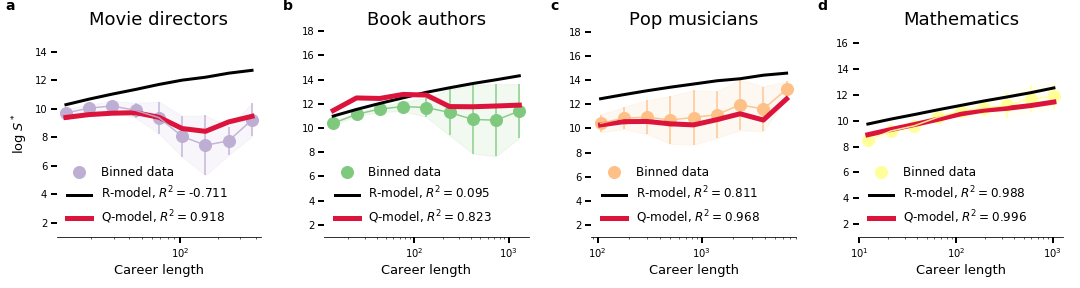}
\caption{{\bf Validation of the Q-model.} We compare the prediction of the {\it Q}-model in terms of the highest impact works of the individuals as the function of their productivity to the random model and to the original data. }
\label{fig:SF9}
\end{figure}

As the random-impact-rule holds, this means that the best product within a creative career occurs at random. However, can we say the same about the magnitude of the success of an individual's best hit? If each artistic product has the same probability to be the most successful, and success does not depend on any intrinsic ability of an individual, success will only be affected by its productivity. this hypothesis   (black lines, Figure \ref{fig:SF9}), known as the \textit{R-model}~\cite{sinatra2016quantifying}, does not capture the observed patterns of impact in artistic domains (colored lines, Figure \ref{fig:SF9}).  This finding was first observed in Ref.~\cite{sinatra2016quantifying} for scientific careers.

We also compared the expected highest impact of the individuals as a function of their productivity based on the $Q$-model. In order to do so we generated synthetic careers by combining the given career length $N_i$ and measured $Q_i$ parameter of the individual $i$, and randomly re-distributed the possible $p_j$ parameters   (picking exactly $N_i$ $p_j$ values for individual $i$) among them to compute the impacts of the synthetic careers by using the Equation proposing the $Q$-model ($S_{i, \alpha} = Q_i p_{i, \alpha}$). After repeating this 100 times to minimize the noise level we arrived at a set of synthetic careers following the $Q$-model. We conducted this comparison on all the studied fields, for which the results are summarized in Table \ref{tab:qval}.

\begin{table}[]
\scriptsize
\begin{tabular}{l|l}
\textbf{Field}                 & \textbf{$R^2$}                   \\
\hline
\hline
Jazz musicians                          & 0.588                            \\
Funk musicians                           & 0.636                            \\
Electronic music artists                        & 0.7                              \\
Movie art directors                   & 0.704266055671                   \\
Script writers                         & 0.808                            \\
Soundtrack composer                       & 0.812                            \\
Hip-hop artists                         & 0.812                            \\
Book authors                          & 0.823                            \\
Classical musicians                     & 0.868                            \\
Rock   musicians                        & 0.871                            \\
Movie directors                       & 0.918                            \\
Movie producers                       & 0.925                            \\
Pop musicians                           & 0.968                            \\
Agronomy                       & 0.985                            \\
Environmental Science         & 0.988                            \\
Biology                        & 0.991                            \\
Space Science or Astronomy  & 0.991                            \\
Zoology                        & 0.992                            \\
Geology                        & 0.993                            \\
Applied Physics               & 0.994                            \\
Engineering                    & 0.994                            \\
Theoretical Computer Science & 0.994                            \\
Chemistry                      & 0.995                            \\
Mathematics                    & 0.996                            \\
Physics                        & 0.998                            \\
Political Science             & 0.999                            \\
Health Science                & 1.0 \\
\hline                               
\end{tabular}
\vspace{0.1cm}
\caption{{\bf Validation of the Q-model.} Goodness of the fit of the {\it Q}-model for the different studied data sets expressed by the measured $R^2$ values.}
\label{tab:qval}
\end{table}






\pagebreak
\newpage

\subsection{Randomness in networking}
\label{sectionSI:network}

\begin{figure}
\centering
\includegraphics[width=1.0\textwidth]{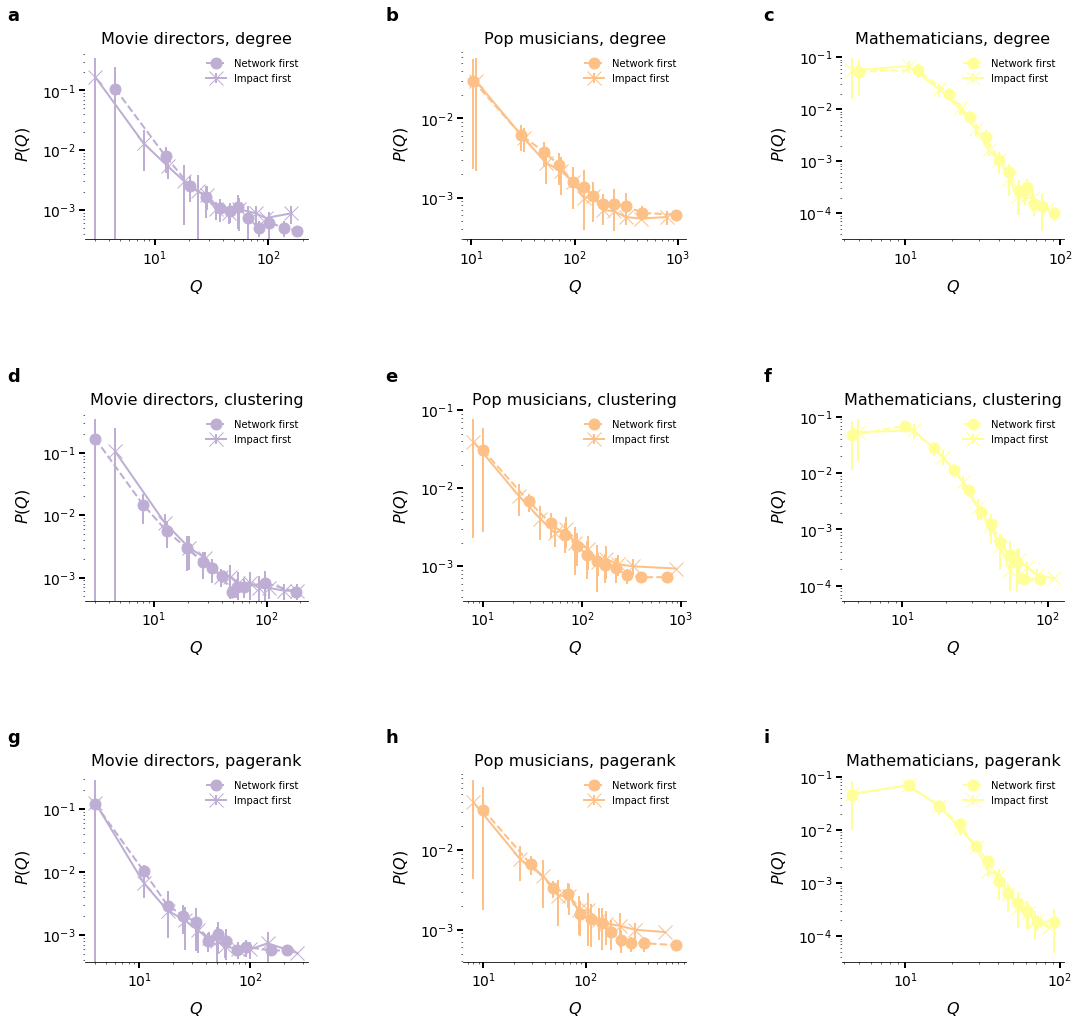}
\caption{\bf $Q$ distributions, where individuals are split based on their professions (director, pop musician, mathematician), while their network centralities are captured by their degree, PageRank, and clustering.}
\label{fig:SF9a}
\end{figure}

\begin{figure}
\centering
\includegraphics[width=1.00\textwidth]{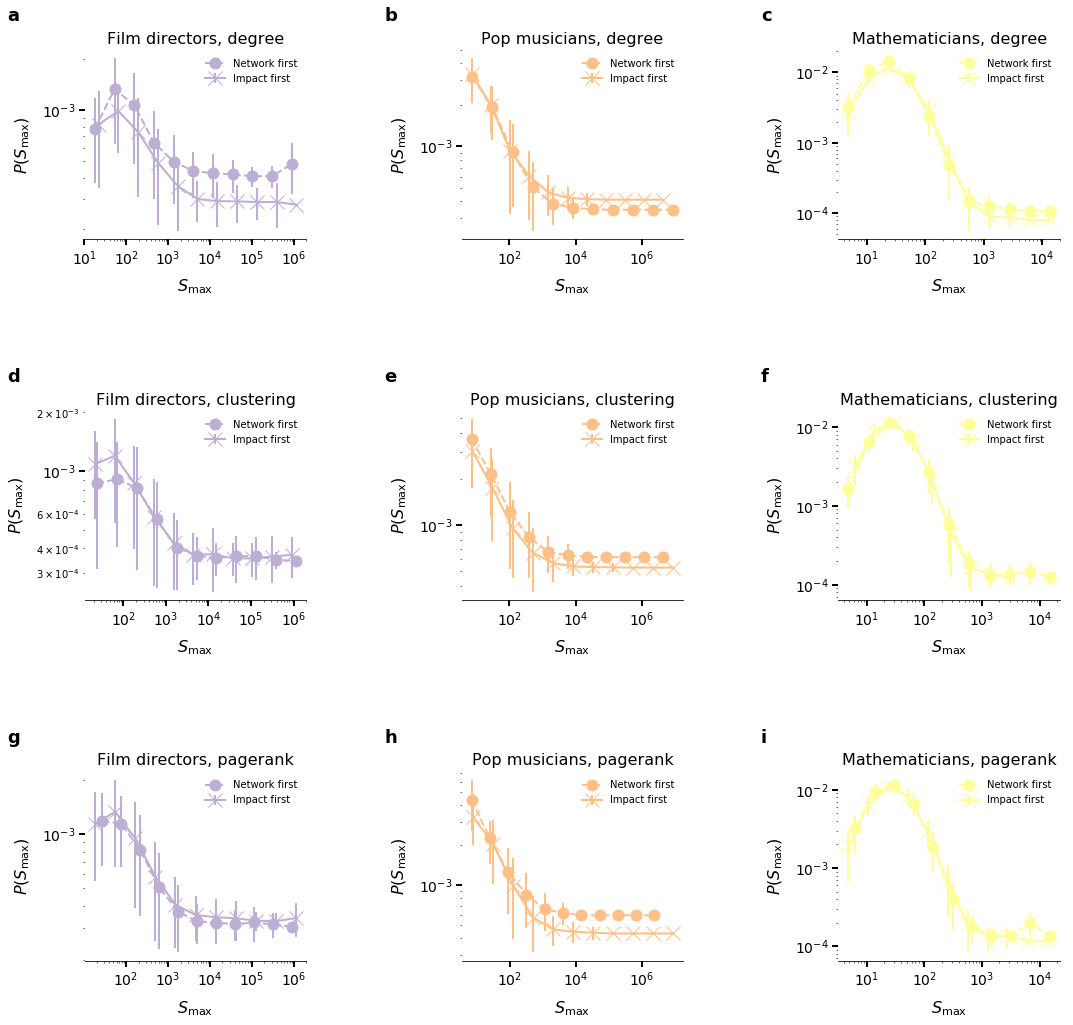}
\caption{\bf $S$ distributions, where individuals are split based on their professions (director, pop musician, mathematician), while their network centralities are captured by their degree, PageRank, and clustering.}\label{fig:SF9b}
\end{figure}

\begin{figure}
\centering
\includegraphics[width=1.00\textwidth]{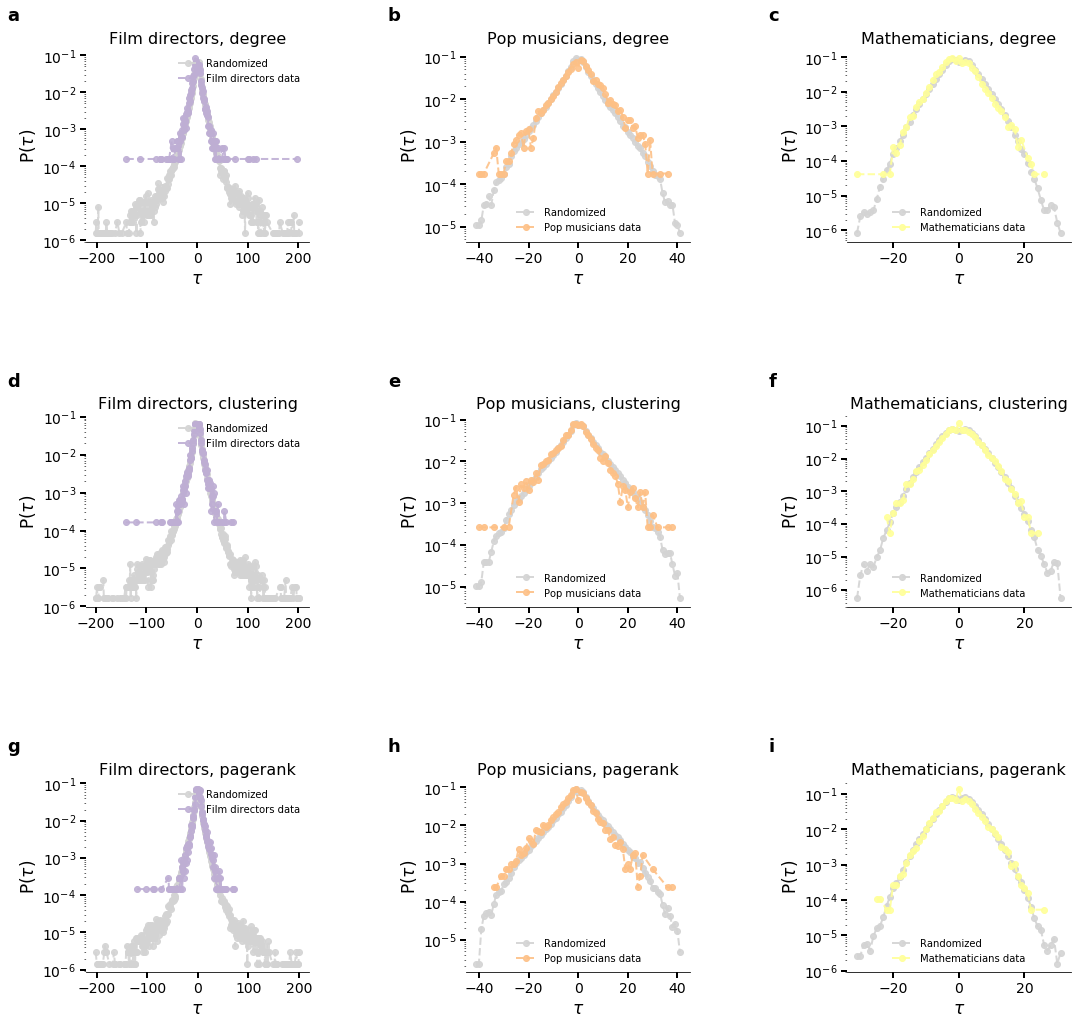}
\caption{\bf Comparing the $\tau$ distributions between the randomized and the measured network data for different careers and network measures.}
\label{fig:SF999}
\end{figure}

We tested the relationship between the collaboration network of an individual and her success for a number of creative fields (movie directors, pop musicians, mathematicians). Results show two different types of networking behavior. For one type of individuals, their impact peaks first, and an increase in network centrality follows. For the others, the opposite is observed. Figure \ref{fig:SF9a}-\ref{fig:SF9b} shows the distribution of the $Q$ parameter and the impact $S$ for these two groups of individuals, their network relevance measured by a set of standard network features, namely  their degree, clustering, and PageRank centrality in the collaboration networks. Results show that there is no significant difference between the success patterns of these two groups (KS statistics in Table \ref{tab:QdistrKSvalues1}-\ref{tab:QdistrKSvalues2}). In Figure \ref{fig:SF999}  and Table \ref{tab:TaudistrKSvalues} We also compare the value of $\tau$, a shifting parameters determined from the data associated to each individuals' career, to the $\tau$ values we obtain in  a randomized null-model data which we generate by reshuffling the original time series.

\begin{table}[]
\scriptsize
\begin{tabular}{l||ll|ll|ll}
           & \multicolumn{2}{c}{Pop music} & \multicolumn{2}{c}{Mathematicians} & \multicolumn{2}{c}{Film directors} \\
           \hline
           \hline
           & $d$           & $p$           & $d$                & $p$           & $d$              & $p$             \\
PageRank   & 0.04769       & 0.10385       & 0.0222             & 0             & 0.02202          & 0.0561          \\
Degree     & 0.03212       & 0.3186        & 0.07452            & 0             & 0.05878          & 0               \\
Clustering & 0.07297       & 0.0028        & 0.05425            & 0             & 0.00662          & 0.99665    \\
\hline
\end{tabular}
\vspace{0.1cm}
\caption{{\bf Results of the Kolmogorov-Smirnov test between the $Q$ distributions. } Individuals are split based on kind of career (director, pop musician, mathematician), while their network centralities are captured by their degree, PageRank, and clustering. The table reports both the measured KS distance values ($d$) and their statistical significance ($p$).}
\label{tab:QdistrKSvalues1}
\end{table}


\begin{table}[]
\scriptsize
\begin{tabular}{l||ll|ll|ll}
           & \multicolumn{2}{c}{Pop music} & \multicolumn{2}{c}{Mathematicians} & \multicolumn{2}{c}{Film directors} \\
           \hline
           \hline
           & $d$          & $p$            & $d$            & $p$               & $d$               & $p$            \\
PageRank   & 0.074        & 4.00E-05       & 0.039          & 1.00E-05          & 0.041             & 0              \\
Degree     & 0.029        & 0.22581        & 0.085          & 0                 & 0.068             & 0              \\
Clustering & 0.073        & 0.00016        & 0.029          & 0.00179           & 0.057             & 0      \\
\hline
\end{tabular}
\vspace{0.1cm}
\caption{{\bf Results of the Kolmogorov-Smirnov test between the $S$ distributions. } Individuals are split based on kind of career (director, pop musician, mathematician). We used the following network measures: degree, PageRank centrality, and clustering.}
\label{tab:QdistrKSvalues2}
\end{table}

\label{subsectionSI:networkrandom}

\begin{table}[]
\scriptsize
\begin{tabular}{l||ll|ll|ll}
           & \multicolumn{2}{c}{Pop music} & \multicolumn{2}{c}{Mathematicians} & \multicolumn{2}{c}{Film directors} \\
           \hline
           \hline
           & $d$           & $p$           & $d$             & $p$              & $d$              & $p$             \\
PageRank   & 0.12928       & 0.55792       & 0.27337         & 1.00E-05         & 0.14286          & 0.60065         \\
Degree     & 0.12978       & 0.51578       & 0.26937         & 1.00E-05         & 0.12158          & 0.79516         \\
Clustering & 0.11455       & 0.69881       & 0.29273         & 0                & 0.15873          & 0.47926       \\
\hline
\end{tabular}
\vspace{0.1cm}
\caption{{\bf Results of the Kolmogorov-Smirnov test between the randomized and the measured network parameters.} The network positions are captured by degree, PageRank, and clustering. We took the fields of pop music, mathematics, and movie directors. The distributions of the corresponding values can be found on Figure \ref{fig:SF999}}.
\label{tab:TaudistrKSvalues}
\end{table}

\clearpage